\documentclass[range,letterpaper]{ar2emod}

\usepackage{epsfig,color}

\def\tlambda{{\tilde\lambda}}
\def\spa#1.#2{\left\langle#1\,#2\right\rangle}
\def\spb#1.#2{\left[#1\,#2\right]}
\def\spab#1.#2.#3{\sandmm#1.#2.#3}
\def\spba#1.#2.#3{\sandpp#1.#2.#3}
\def\spaa#1.#2.#3.#4{\sandmp#1.{#2#3}.#4}
\def\spbb#1.#2.#3.#4{\sandpm#1.{#2#3}.#4}
\def\spash#1.#2{\vphantom{\hat K}\spa{\smash{#1}}.{\smash{#2}}}
\def\spbsh#1.#2{\vphantom{\hat K}\spb{\smash{#1}}.{\smash{#2}}}
\def\lor#1.#2{\left(#1\,#2\right)}
\def\sand#1.#2.#3{%
\left\langle\smash{#1}{\vphantom1}^{-}\right|{#2}%
\left|\smash{#3}{\vphantom1}^{-}\right\rangle}
\def\sandp#1.#2.#3{%
\left\langle\smash{#1}{\vphantom1}^{-}\right|{#2}%
\left|\smash{#3}{\vphantom1}^{+}\right\rangle}
\def\sandpp#1.#2.#3{%
\left\langle\smash{#1}{\vphantom1}^{+}\right|{#2}%
\left|\smash{#3}{\vphantom1}^{+}\right\rangle}
\def\sandpm#1.#2.#3{%
\left\langle\smash{#1}{\vphantom1}^{+}\right|{#2}%
\left|\smash{#3}{\vphantom1}^{-}\right\rangle}
\def\sandmp#1.#2.#3{%
\left\langle\smash{#1}{\vphantom1}^{-}\right|{#2}%
\left|\smash{#3}{\vphantom1}^{+}\right\rangle}
\def\sandmm#1.#2.#3{%
\left\langle\smash{#1}{\vphantom1}^{-}\right|{#2}%
\left|\smash{#3}{\vphantom1}^{-}\right\rangle}
\def\spab#1.#2.#3{\sandmm#1.#2.#3}
\def\spbb#1.#2.#3.#4{\sandpm#1.{#2#3}.#4}
\newbox\charbox
\newbox\slabox
\def\s#1{{      
        \setbox\charbox=\hbox{$#1$}
        \setbox\slabox=\hbox{$/$}
        \dimen\charbox=\ht\slabox
        \advance\dimen\charbox by -\dp\slabox
        \advance\dimen\charbox by -\ht\charbox
        \advance\dimen\charbox by \dp\charbox
        \divide\dimen\charbox by 2
        \raise-\dimen\charbox\hbox to \wd\charbox{\hss/\hss}
        \llap{$#1$}
}}
\def\ksl{\s{k}}
\def\Ksl{\s{K}}
\def\ssl{\s{s}}
\def\kflat{ {k^\flat} }

\def\Shift#1#2{{[#1,#2\rangle}}
\def\Res{\mathop{\rm Res}}
\def\oneloop{{1 \mbox{-} \rm loop}}

\def\InfPart#1#2{\mathop{\rm Inf}\limits_{#1}{#2}}

\newcommand{\Bmp}[1]{\langle #1\rangle}
\newcommand{\Kf}[1]{\tilde K_{#1}}

\newcommand{\Kfm}[1]{\tilde K^{-}_{#1}}

\begin{document}

\input epsf.tex    
\input epsf.def   

\input psfig.sty

\jname{Ann. Rev. Nucl. Part. Sci.} \jyear{2010} \jvol{}
\ARinfo{\hfill MIT-CTP 4092 \,\,\, CERN-PH-TH/2009-255 \,\,\,
NIKHEF/2009-035}

\title{\mbox{ } \vspace*{-20mm} \\ Multi-Parton Scattering Amplitudes via On-Shell Methods}

\markboth{Multi-Parton Scattering Amplitudes via On-Shell Methods}{Multi-Parton
Scattering Amplitudes via On-Shell Methods}

\author{Carola F. Berger
\affiliation{Center for Theoretical Physics, Massachusetts Institute of Technology,
Cambridge, MA 02139, USA}
Darren Forde
\affiliation{Theory Division, CERN, CH-1211 Geneva 23, Switzerland, \\
NIKHEF Theory Group, Science Park 105, NL-1098 XG Amsterdam, The Netherlands} }


\begin{abstract}
We present an overview of recent developments, based on on-shell techniques, in the calculation of
multi-parton scattering amplitudes at one loop that are relevant for phenomenological
studies at hadron colliders. These new on-shell methods make efficient use of
the physical properties of the hard scattering, such as unitarity and
factorization.
\end{abstract}


\maketitle
\newpage

\section{Introduction}

With the recent first collisions at the LHC we are entering a new
era of discovery in particle physics. Colliding protons at very
high energies makes the LHC a fertile environment for the
production of high-multiplicity events. If we are to take full
advantage of the discovery potential at high energy hadron
colliders such as the Tevatron and the LHC, we need to have a
precise understanding of the physics that will occur there. This
necessitates computations of the Standard Model background,
especially of QCD processes, to at least next-to-leading order
(NLO) in the perturbative series.

Historically, the bottleneck in NLO computations has been the
one-loop virtual contributions. Over the last few years rapid
progress has been made in the development of new techniques for
these one-loop computations. These advancements have been
motivated both by a greater desire to understand the structure of
scattering amplitudes as well as  the need for improved efficiency
and automation in the computation of one-loop matrix elements.
These matrix elements are needed, for example, for the precise computation of
many background processes at the LHC (see
e.\,g.~\cite{Bern:2008ef}). The ability to automate and
``mass-produce'' amplitudes requires approaches that are both
numerically stable and straightforward to implement as an
algorithm.

The standard approach to performing this class of computations has
heavily relied upon Feynman diagram techniques. There have been
many impressive results with this approach (see \cite{Bern:2008ef}
and references therein; for example, cross sections for 6-point
processes that have been computed via Feynman diagrammatic methods
include~\cite{Feynman6}). However, Feynman diagrams suffer from
two problems. For one, there is a factorial growth in the number
of terms as the multiplicity of partons in the process increases.
Furthermore, each Feynman diagram is gauge dependent. This means
that there will be large cancellations between terms that combine
to give the gauge-independent amplitudes. It is these two problems
that make automated approaches using Feynman diagrams difficult as
the number of partons increases.

Consequentially, the focus of much recent progress has been to side-step these issues.
On-shell recursion and unitarity methods work with gauge-independent amplitudes as building blocks instead of Feynman diagrams.
It is these new techniques that we focus on in this review. An earlier review of these techniques was presented in~\cite{Bern:2007dw}.
However, many of the details presented there have been superseded by even more efficient techniques which we present in the following.

In general, as we will explain in more detail below, a one-loop
amplitude can be decomposed into a set of scalar box, triangle,
bubble, and tadpole integrals, that is, integrals with four,
three, two, or one loop propagators, respectively. These scalar
integrals contain all the logarithmic and polylogarithmic
dependence of the amplitude and are multiplied by rational
coefficients. In addition, there are also purely rational terms in
the amplitude.

The original unitarity approach of \cite{Bern:1994cg} applied
two-particle cuts in four dimensions and was used to produce many
results~\cite{Bern:1997sc}. The terms containing (poly)logarithms
and associated constants could all be computed via two-particle
cuts. Terms with only a rational dependence on momentum invariants
however
 could not be computed in this way and separate techniques were required~\cite{Bern:1997sc}.
A systematic approach
for the computation of these rational terms without the use of Feynman diagrams did not appear until
a few years ago. The starting point for these developments were the recursion relations,
developed at tree level by Britto et al. (BCFW)~\cite{BCFW}. An earlier version of a tree-level recursion relation was the
off-shell Berends-Giele recursion~\cite{Berends:1987me}.
The proof of the BCFW on-shell recursion relations only relies upon the factorization
properties of the amplitudes and on Cauchy's theorem. They could therefore be adapted to the more
complicated problem of computing the rational parts of loop amplitudes~\cite{Bern:2005cq,Berger:2006ci}.
On-shell recursion for the rational parts of loop amplitudes was first used in an
analytic context~\cite{Berger:2006ci} and then
further adapted into a numerical procedure~\cite{Berger:2008sj}.

At the same time, improvements to the original unitarity approach
were also occurring. Brandhuber et al. used two-particle cuts with
MHV vertices to compute certain sets of one-loop
amplitudes~\cite{Brandhuber}. The application by Britto et al. of
generalized unitarity~\cite{Bern:1997sc,Britto:2004nc} to the
computation of box coefficients highlighted the benefits of
examining not the one-loop integral as a whole, but its integrand.
The work of Ossola, Papadopoulos and Pittau
(OPP)~\cite{Ossola:2006us,Mastrolia:2008jb} followed in this vein.
Upon separation of the integrand into a standard set of basis
terms, the problem could be reduced to solving for their
coefficients numerically. The simple analytic extraction of
triangle and bubble coefficients was the focus of the work by one
of the authors~\cite{Forde:2007mi}, where the known analytic
behavior of the integrand was used to straightforwardly extract
the bubble and triangle coefficients. A numerical adaptation of
this procedure suitable for automation was then presented in
Ref.~\cite{Berger:2008sj}.

The investigation of the one-loop integral itself has also presented new directions to explore.
Britto et al.~\cite{Britto:2004nj} showed how the integral when
written in a canonical form can be directly integrated. This procedure was developed further
to produce new results~\cite{Britto:2008vq} and also to provide a general analytic structure
for the coefficients of one-loop amplitudes~\cite{Britto:2007tt,Britto:2009wz}.
Taking further advantage of the
analytic properties of the two-particle cut amplitude, Mastrolia~\cite{Mastrolia:2009rk} applied Stokes'
theorem and a generalized residue theorem to compute bubble coefficients via direct integration.

Extending the four-dimensional cut techniques to $D$ dimensions
(in dimensional regularization) has also provided a second very
fruitful approach for the computation of the rational terms. In
$D$ dimensions the rational terms develop branch cuts and are so
accessible via unitarity cuts. Giele et
al.~\cite{Giele:2008ve} elucidated the extra
structures present in $D$ dimensions beyond those of the original
four-dimensional OPP integrand. Hence they were able to compute
the rational terms. Following up on this and an earlier result
relating masses to $D$-dimensional unitarity~\cite{Bern:1995db},
Badger presented an alternative computational approach where the
$(D-4)$-dimensional terms are treated as an additional mass in the
loop~\cite{Badger:2008cm}. A different approach advanced by
Papadopoulos et al.~\cite{OPPrat} involves splitting the rational
computation into two pieces. One part is computed from the
one-loop integrand and is an extension to the original OPP
approach, while the second part comes from a reduced form of
Feynman diagrams.

These approaches have been implemented in several automated tools for the computation of one-loop amplitudes:
\texttt{BlackHat}~\cite{Berger:2008sj},
\texttt{CutTools/OneLOop}~\cite{Ossola:2007ax,vanHameren:2009dr}, \texttt{Rocket}~\cite{Ellis:2008qc}, and others~\cite{othercodes}.
These programs, combined with tools for the real
emission part~\cite{Gleisberg:2008ta, Frederix:2009yq} have yielded a host of new
results at next-to-leading order~\cite{Berger:2009zg,KeithEllis:2009bu,Melnikov:2009dn,Ossola:2007bb,Binoth:2008kt,Bevilacqua:2009zn}.

Below we will review the main ideas of all these developments. Due to lack of space, we regret that we can neither
present all details nor an exhaustive list of all results and refer the reader to the cited literature and references therein.

We begin with a review of our notation and the general structure
of multi-parton scattering amplitudes and explain how to construct
such amplitudes recursively at tree level. We will use the spinor
formalism, and those readers unfamiliar with spinor techniques and
their application to the calculation of tree-level and one-loop
amplitudes may wish to consult for example
Refs.~\cite{Mangano:1990by,Dixon:1996wi,Bern:1996je}. Section
\ref{unitarity} discusses the use of (generalized) unitarity to
construct one-loop amplitudes from tree level amplitudes, both the
cut and the rational part. In Section \ref{recursion} we explain
how to alternatively employ a recursive approach to obtain the
purely rational non-logarithmic terms of amplitudes that cannot be
constructed via unitarity in four dimensions. We conclude with a
summary and give an outlook on the expected progress in the near
future.

\section{Structure of Amplitudes}

Below we briefly review our notation, which closely follows Ref.~\cite{Dixon:1996wi} (see also
\cite{Mangano:1990by}).
We then discuss the general structure of tree and one-loop amplitudes in renormalizable gauge theories
to set the stage for the subsequent sections which discuss new methods
for the computation of these amplitudes.

\subsection{Notation and Color Decomposition}

In the following we will consider amplitudes where all
color\footnote{Color here and below refers to any group theory factors.} and coupling information has
been stripped off.
We can express any amplitude
in terms of some basic color (and coupling) factors which are multiplied by
color-ordered subamplitudes, or primitive
amplitudes. These primitive amplitudes, which are defined for specific cyclic orderings
of the external partons,
 carry all the kinematic information but no explicit
color indices.
The full amplitude is then assembled from the primitive amplitudes by dressing them with
appropriate color factors. There exist different, but equivalent, ways of grouping the primitive
amplitudes together into so-called partial amplitudes, and we refer the reader to the
 literature for further information, see \cite{Mangano:1990by,Dixon:1996wi,color,HELAC} and references therein.
In what follows below, we will discuss the computation of the primitive amplitudes,
concentrating on the description of the evaluation of the kinematic part.
As shown recently in Ref.~\cite{Giele:2009ui},
the algorithms that we will present below can be extended to amplitudes with color information.

We express the primitive amplitudes in terms of spinor inner
products,{\footnote{{Note that we use the sign convention of most
of the QCD literature, in the ``twistor'' literature a different
sign convention for $\spb{j}.{l}$ is used, for example in
Refs.~\cite{BCFW}.}}}
\begin{equation}
\spa{j}.{l} = \langle j^- | l^+ \rangle = \bar{u}_-(k_j) u_+(k_l)\,,
\hskip 2 cm
\spb{j}.{l} = \langle j^+ | l^- \rangle = \bar{u}_+(k_j) u_-(k_l)\, ,
\label{spinorproddef}
\end{equation}
where $u_\pm(k)$ is a massless two-component (Weyl) spinor with momentum $k$ and positive
or negative chirality, respectively, which we also write as,
\begin{equation}
 ( \lambda_i )_\alpha \equiv \left[u_+(k_i)\right]_\alpha , \qquad ( \tlambda_i )_{\dot{\alpha}}
\equiv \left[u_-(k_i)\right]_{\dot{\alpha}} \,.
\label{lambdadef}
\end{equation}
Massless four-momenta can be reconstructed from the spinors by,
\begin{equation}
k_i^\mu (\sigma_\mu)_{\alpha \dot{\alpha}} = \left(\ksl_i\right)_{\alpha \dot{\alpha}} =
( \lambda_i )_\alpha  ( \tlambda_i )_{\dot{\alpha}} \, .
\end{equation}
Spinor products can thus be used to construct the usual momentum dot products via
\begin{equation}
\spa{i}.{j} \spb{j}.{i} = \frac{1}{2} \mbox{Tr} \left[ \ksl_i \ksl_j \right] =
2 k_i \cdot k_j = s_{ij} \,.
\end{equation}
Furthermore, we use the following notation for sums of cyclically-consecutive external
momenta and their invariant masses,
\begin{eqnarray}
K_{i\dots j}^\mu & \equiv & k_i^\mu + k_{i+1}^\mu + \cdots + k_{j-1}^\mu + k_j^\mu \, , \\
s_{i \dots j} & \equiv & K_{i\dots j}^2 \, ,
\end{eqnarray}
where all indices are to be understood mod $n$ for $n$-particle amplitudes.

The above formalism can be extended to include massive spinors and vectors, using
the well-known decomposition of any, not necessarily light-like, four-vector $k$ into a sum of
two light-like four-vectors:
\begin{equation}
k^\mu = \kflat^\mu + \frac{k^2}{2 k \cdot q} q^\mu \, .
\end{equation}
Here $q$ is a fixed light-like four-vector, fixing the
axis of the spin for spinors,
and $\kflat$ is the associated projection of the massive vector $k$.
Massive spinors can be constructed via
\begin{eqnarray}
u_{-} (k, q) & = & \frac{1}{\spa{\kflat}.q} \left(\ksl + m \right) | q^+ \rangle =
| \kflat^- \rangle + \frac{m}{\spa{\kflat}.q} | q^+ \rangle \, , \\
u_{+} (k, q) & = & \frac{1}{\spb{\kflat}.q} \left(\ksl + m \right) | q^- \rangle =
| \kflat^+ \rangle + \frac{m}{\spb{\kflat}.q} | q^- \rangle \, ,
\end{eqnarray}
and similarly for the conjugate spinors~\cite{Kleiss:1985yh}.
Here, the label $\pm$ indicates that the spinors $u_{\pm}$ are
eigenstates of the projector $(1 \pm \ssl \gamma^5)$, with the
spin vector $s = k/m - m/(k \cdot q) \, q$. The normalization is
chosen to allow a smooth limit to the massless case.

\subsection{On-Shell Recursions at Tree Level}
\label{treerecursion}

An efficient recursive technique for computing tree-level multi-parton scattering amplitudes
was developed more than 20 years ago~\cite{Berends:1987me} and adapted for numerical implementation
in various
computer codes~\cite{HELAC,ALPGEN,COMIX}.  Berends-Giele recursion connects smaller off-shell currents together to produce amplitudes.
More recently it was realized that through the use of complex kinematics amplitudes
can be computed entirely using only smaller on-shell amplitudes. This leads to more compact analytic expressions not only at tree level but for rational terms also at loop level.
Here we briefly review the on-shell recursion
relations for tree level amplitudes found and proved in
Refs.~\cite{BCFW}. A recursive approach at loop level is not quite
so straightforward, as we will discuss in Section~\ref{recursion}.

At tree level, the on-shell recursion relations rely on general properties
of complex functions as well as on factorization properties of
scattering amplitudes.
The proof~\cite{BCFW} of the tree-level relations employs a
parameter-dependent complex continuation ``$[j,l\rangle$'', or
``shift'', of two of the external massless spinors, $j$ and $l$,
in an $n$-point process,
\begin{equation}
\Shift{j}{l}:\hskip 2 cm
\tlambda_j \rightarrow \tlambda_j - z\tlambda_l \,,
\hskip 2 cm
\lambda_l \rightarrow \lambda_l + z\lambda_j \,.
\label{SpinorShift}
\end{equation}
where $z$ is a complex number.  The corresponding momenta
are then continued in the complex plane as well,
whereby they remain massless, $k_j^2(z) = 0 = k_l^2(z)$,
and overall momentum conservation is maintained.

An on-shell amplitude containing the momenta $k_j$ and $k_l$
then also becomes parameter-dependent,
$A(z)$. The physical amplitude is given by $A(0)$.
When $A$ is a tree amplitude or finite one-loop
amplitude, $A(z)$ is a rational function of $z$.
At tree level, $A(z)$ only has simple poles. These poles arise only from the
 shifted propagators of the amplitude. For example,
\begin{equation}
\frac{i}{K_{r \dots l \dots s}^2} \rightarrow  \frac{i}{K_{r \dots s}^2 + z
\sand{j}.{\Ksl_{r\ldots s}}.{l} } \, , \label{propagator}
\end{equation}
if the set of legs $\{r, \dots ,s\}$ contains leg $l$,
which is shifted according
to eq.~(\ref{SpinorShift}).
In the vicinity of the location of the pole $z_{rs}$, the complex continued
amplitude is then  schematically given by,
\begin{equation}
\lim\limits_{z \rightarrow z_{rs}} A(z) =
\sum_{h} A^h_{L}(z) {i \over
K_{r \dots s}^2 + z
\sand{j}.{\Ksl_{r\ldots s}}.{l}  } A^{-h}_R(z) \, ,
\end{equation}
where $h=\pm1$ labels the helicity of the intermediate state, and
the labels $L$ and $R$ denote amplitudes with fewer legs, which
the propagator eq.~(\ref{propagator}) connects.
The number of poles $z_{rs}$ in the complex plane is given by the number of ways
the set of external legs can be partitioned such that the legs $j$ and $l$
always appear on opposite sides of the $z$-dependent
propagator.

\begin{figure}[hp]
\centerline{\epsfig{file=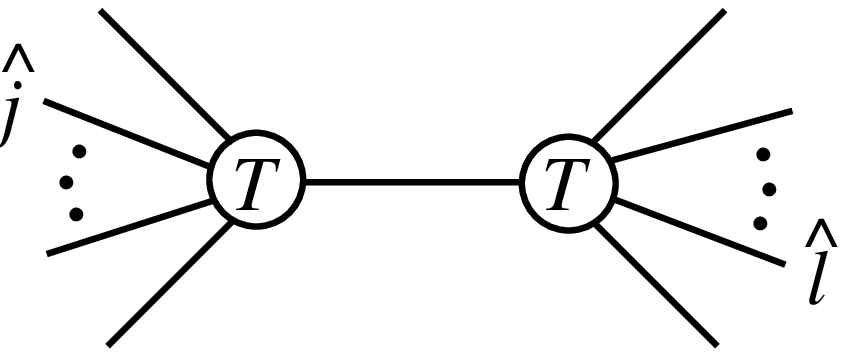,height=2.5cm}}
\caption{Schematic representation of a typical tree recursive
contribution to eq.~(\ref{BCFW}).  The labels `$T$' refer to
on-shell tree amplitudes. The momenta $\hat{j}$ and $\hat{l}$
are complex continued, on-shell momenta. }
\label{TreeGenericFigure}
\end{figure}

We can now use Cauchy's theorem,
\begin{equation}
{1\over 2\pi i} \oint_C {dz\over z}\,A(z)  = 0\,,
\label{ContourInt}
\end{equation}
where the contour $C$ is taken around the circle at infinity,
and the integral vanishes if the
complex continued amplitude $A(z)$ vanishes as $z \rightarrow \infty$.
Evaluating the integral as a sum
of residues, we can then solve for the physical amplitude $A(0)$ to obtain,
\begin{equation}
A(0) = -\sum_{{\rm poles}\ \alpha} \Res_{z=z_\alpha}  {A(z)\over z}\, =
\sum_{r,s} \sum_{h}
    A^h_{L}(z = z_{rs}) { i \over K_{r\cdots s}^2 } A^{-h}_R(z = z_
{rs})  \,.
\label{BCFW}
\end{equation}
The on-shell amplitudes with fewer legs, $A_{L}$ and $A_R$,
are evaluated in kinematics that have been shifted by eq.~(\ref{SpinorShift}) with
$z=z_{rs}$, where eq.~(\ref{propagator}) has a pole,
\begin{equation}
z_{rs} = - {K_{r\cdots s}^2 \over \sand{j}.{\Ksl_{r\ldots s}}.{l} } \,.
\end{equation}
In the following, such shifted, on-shell momenta will be denoted by
$k(z = z_{rs}) \equiv \hat{k}$.
A typical contribution to the sums in eq.~(\ref{BCFW}) is
illustrated in Fig.~\ref{TreeGenericFigure}.

We have thus succeeded in expressing the $n$-point amplitude $A$ in terms
of sums over on-shell, but complex continued, amplitudes with fewer legs,
which are connected by scalar propagators.
These recursion relations can be extended to massive
QCD and other
theories~\cite{masses,gravity}.
Moreover, for certain helicity
configurations, this recursion relation can be solved explicitly, leading
to new all-multiplicity expressions for these amplitudes~\cite{allorder}.

The basic ingredients to obtain such a recursion relation are
complex momenta and analysis, which are necessary to make 3-point
vertices non-vanishing; factorizability, which is responsible for
the simple pole structure; and the vanishing of the boundary
contribution as $z \rightarrow \infty$. At tree level in QCD, one
can always find complex continuations where this boundary
condition vanishes. However, as we will see in
Section~\ref{recursion} below, this is not the case at the
one-loop level, and a recursive approach becomes considerably more
complicated. Other theories such as a scalar $\phi^4$ theory have
non-vanishing $z \rightarrow \infty$ behavior already at the tree
level, which spoils the recursive approach. Studies of the origin
of these boundary terms and their relation to the Lagrangian can
be found in Refs.~\cite{boundary}. Furthermore, additional
amplitude structures related to on-shell recursion relations and
twistor space have been uncovered~\cite{amplitudestructure}.

\subsection{Structure of One-Loop Amplitudes}

We now turn to the discussion of gauge-theory one-loop amplitudes,
the main subject of this review.
From here on, we will denote one-loop amplitudes explicitly with a superscript
$A^{\oneloop}$, and tree amplitudes without superscript simply by $A$.

\begin{figure}[hp]
\centerline{\epsfig{file=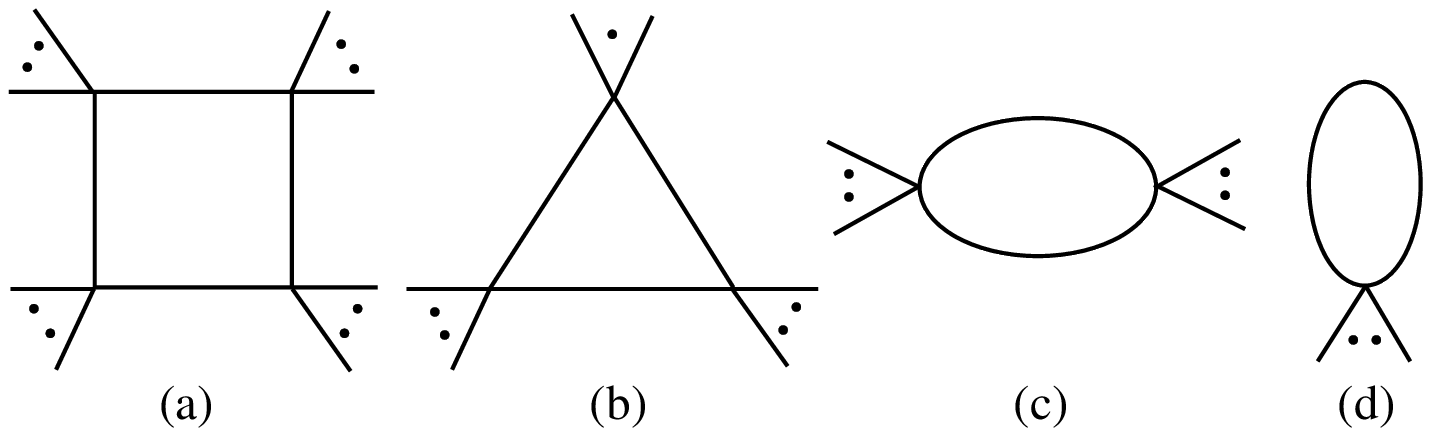,height=4cm}}
\caption{Representative examples of integrals appearing
in eq.~(\ref{integralbasis}): (a) a box
integral, that is, a 4-point integral,
(b) a triangle (3-point) integral, (c) a bubble (2-point) integral, and
(d) a tadpole integral. Each corner can have one or more
external momenta attached to it. The tadpole vanishes when all
internal propagators are massless. }
\label{Basisfig}
\end{figure}

Using reduction techniques~\cite{reduction,vanNeerven:1983vr,dimshift}
any $m$-point scalar integral, $m > 4$, can be reduced to scalar integrals
with at most four propagators. That is, any one-loop
 $n$-point amplitude $A_n^{\oneloop}$ can be decomposed into a basis $B_4$ of
scalar box, triangle, bubble and tadpole integrals, with rational
coefficients in four dimensions. In $D$ dimensions, for example
when working in dimensional regularization where $D = 4 - 2
\varepsilon$, then the basis $B_D$ is extended to include a scalar
pentagon. The  coefficients of the $D$-dimensional basis scalar
integrals  can be decomposed into purely four-dimensional
coefficients after expanding in $\varepsilon$. Purely rational
terms are generated when terms higher order in $\varepsilon$ in
the coefficients are multiplied by the poles in the integrals,
\begin{equation}
A_n^{\oneloop} = \sum\limits_{j \in B_D} c^D_j
\mathcal{I}^D_j = \sum\limits_{j \in B_4} c^{D=4}_j \mathcal{I}^D_j
+ {\cal R}_n \label{integralbasis} \,.
\end{equation}
Illustrative examples of integrals of the basis $B$ are shown in Fig.~\ref{Basisfig}.

We will see in the next two sections how to obtain these
coefficients and rational terms in efficient ways. The scalar
integrals contain infrared and ultraviolet divergences that are
regulated via dimensional regularization, and depend
logarithmically or polylogarithmically on momentum-invariants. The
integrals appearing in eq.~(\ref{integralbasis}) are known and
tabulated for example in Refs.~\cite{Bern:1994cg,Ellis:2007qk}.
 In order to compute one-loop matrix elements the task is therefore
reduced to the determination of the coefficients, it is not necessary to perform
any integrals.

 For amplitudes with only massless particles, the tadpoles
vanish. In $\mathcal{N} = 4$ supersymmetric theories, as counting
of powers of loop momenta in one-loop integrals reveals, only box
integrals contribute with $(D = 4)$-dimensional coefficients
(i.\,e. free of $\varepsilon$ terms), and in $\mathcal{N} = 1$
supersymmetry bubble, triangle, and box integrals contribute, with
four-dimensional coefficients~\cite{Bern:1994cg}. That is,
theories with unbroken supersymmetries do not contain purely
rational terms that are not associated with any of the integrals
in the basis. One-loop supersymmetric amplitudes can therefore be
completely reconstructed from unitarity cuts, as we will now
discuss.

\section{Extraction of Integral Coefficients via Unitarity}
\label{unitarity}

The problem of computing eq.~(\ref{integralbasis}) has been reduced to determining the coefficients, $c_j$, multiplying the known basis integral functions, in the most efficient manner possible. The nature of eq.~(\ref{integralbasis}) suggests the use of unitarity cuts to isolate particular integral coefficients.  At the most basic level a unitary cut effectively replaces a propagator with an on-shell delta function, i.e. we ``cut'' the propagator with the replacement,
\begin{eqnarray}
\frac{1}{p^2-m^2 +  i\epsilon}\rightarrow \delta^{(+)}(p^2-m^2).
\end{eqnarray}
In the original unitarity approach~\cite{Bern:1994cg} only two propagators were cut but more recent developments have highlighted the benefits of applying multiple cuts \cite{Bern:1997sc,Britto:2004nc}. The application of multiple cuts is known as Generalized Unitarity and has become the foundation of the most recent developments in the literature.

Our starting point is the form of eq.~(\ref{integralbasis})
decomposed in the $B_4$ basis,  we postpone the discussion of the
general $D$-dimensional case using $B_D$ to
Section~\ref{section:D-Dimensional_Generalized_Unitarity}. The
purely rational terms are independent of any possible cuts in four
dimensions and therefore only the remaining ``cut-constructible''
pieces are accessible via a unitarity technique.

We apply a number of cuts to the expression of the one-loop amplitude and match this expression to that of the basis decomposition, eq.~(\ref{integralbasis}), with the same set of cuts applied. This allows us to directly relate the cut expression to the basis integral coefficients. This procedure is repeated with as many different sets of cuts as is needed to compute all the basis coefficients. Rather than actually applying the cuts to the full expression for the one loop amplitude, computed for example with Feynman diagrams, we construct the cut expression simply by multiplying appropriate on-shell tree amplitudes together.  This allows us to take advantage of efficient, compact forms of tree amplitudes produced via recursion relations, for example those of Section~\ref{treerecursion}.

Below, we describe two unitarity approaches for the extraction of integral coefficients. They both utilize
knowledge of the {\it integrand}, $\tilde{A}(l)$, of the one-loop amplitude, $A_n^{\oneloop} =\int dl \tilde{A}_n(l)$,
to derive the basis integral coefficients. The first, described in Section~\ref{section:genuni}, is based upon the
examination of the behavior of the loop integrand and the loop momenta in the complex plane. The second, described
in Section~\ref{section:OPP}, known as the Ossola, Papadopoulos and Pittau (OPP) method, relies upon computing the
coefficients of the general structure of the loop integrand itself.

\subsection{Generalized Unitarity in Four Dimensions}
\label{section:genuni}

In general we want to isolate as few basis integral coefficients as possible with each cut that we consider.
It is easy to see that a quadruple cut can be used to isolate,  on the basis side, a single box coefficient.
There are not enough propagators in the bubble and triangle integrals to accommodate so many cuts. The set of
cuts we require to compute all box coefficients corresponds simply to all possible boxes that could be present.

\subsubsection{Boxes}

 A very straightforward way to extract a specific box coefficient from a quadruple cut expression was proposed by Britto et al. in~\cite{Britto:2004nc}. The momentum circulating inside a loop without cuts is off-shell and can therefore be parametrized in terms of four free components. Applying a cut to one of the propagators in the expression of the one-loop amplitude reduces the number of free components in this loop momentum by one.  Second, third and fourth cuts will then reduce the number of free components to zero. The loop momentum for the box is then completely frozen by the four delta-function constraints. The desired scalar box coefficient can now be read off from the resulting rational expression,
\begin{equation}
\int \frac{d^4 l}{(2 \pi)^4} \left.\frac{ d(l)}{(l^2 + i \epsilon) (l_1^2 + i \epsilon)
(l_2^2 + i \epsilon)(l_3^2 + i \epsilon)}\right|_{l_i^2 + i \epsilon \rightarrow
\delta^+ (l_i^2)} \longrightarrow \frac{1}{2} \sum\limits_{\mbox{\tiny solutions}} d(l_{\mbox{\tiny solution}}) \, .
\end{equation}

\begin{figure}[hp]
\centerline{\epsfig{file=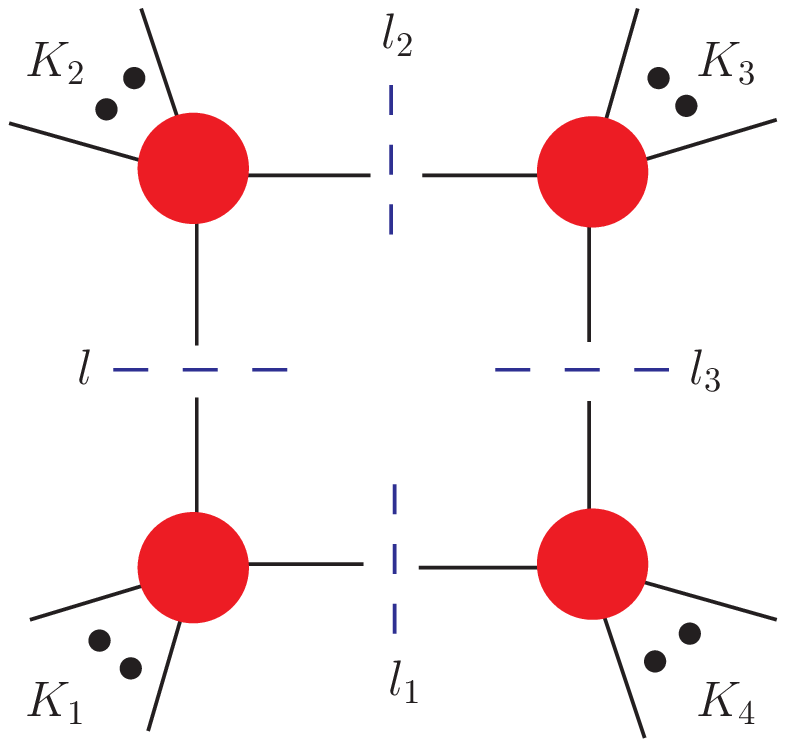,height=6cm}}
\caption{A quadruple cut one-loop integral isolating the single box term $d_0(K_1^2,K_2^2,K_3^2,K_4^2)$. }
\label{Box_coeff}
\end{figure}

The coefficient is a product of four tree amplitudes that sit at
the four corners of the cut box as illustrated in
Fig.~\ref{Box_coeff}. The momenta flowing into the trees satisfy
the four cut constraints.  In general there are two possible
solutions to the constraints parameterizing the box.  For example
with at least one massless leg (leg 1) we
have~\cite{Berger:2008sj,Risager:2008yz},
\begin{eqnarray}
l^{\mu}_{\pm}&=&\frac{\Bmp{1^{\mp}|\s K_2\s K_3 \s K_4 \gamma^{\mu}|1^{\pm}}}{2\Bmp{1^{\mp}|\s K_2\s K_4|1^{\pm}}}.\label{eq:box_solutions}
\end{eqnarray}
Inserting each solution into the product of four trees at each corner and then summing the two results gives the complete box coefficient,
\begin{eqnarray}
d_{0}=\frac{1}{2}\sum_{a=\pm}d_{a},\;\;\;\;\;\;\;\;\;\; d_a=A_1(l_{a})A_2(l_{a})A_3(l_{a})A_4(l_{a}).\label{eq:box_coeffs}
\end{eqnarray}
Further discussion on the use of both solutions and alternative approaches can be found in~\cite{Cachazo:2008vp}.

\subsubsection{Triangles}

To compute triangle coefficients, we first apply
 a triple cut to our one-loop amplitude. Unlike the box case, this cut isolates not just a
 single triangle but also any boxes which also contain the same triple cut.
 Furthermore we are left with a loop integral with a single free component.
 Our cut expression therefore still depends upon the loop integration. To extract the
 triangle coefficient isolated by the triple cut therefore requires two steps. First,
 we must remove the boxes polluting the triple cut expression. Then we must relate the
 triangle integral, which depends upon the remaining free loop-parameter, to the scalar triangle basis integral.

As was first proposed in~\cite{Forde:2007mi}, both issues can be solved simultaneously by examining the analytic behavior of the triple cut expression in the free parameter $t$ of the loop momentum. We choose the following specific parametrization of the loop momentum,
\begin{eqnarray}
l^{\mu}=K_1^{\flat,\mu}+K_2^{\flat,\mu}+\frac{t}{2}\Bmp{K^{\flat,-}_1|\gamma^{\mu}|K^{\flat,-}_2}
+\frac{1}{2t}\Bmp{K^{\flat,-}_2|\gamma^{\mu}|K^{\flat,-}_1}.\label{eq:sol_for_l_in_3_mass}
\end{eqnarray}
The massless momenta $K_i^{\flat,\mu}$ are given by
\begin{eqnarray}
&K_1^{\flat,\mu}=\gamma\alpha\frac{\gamma K^{\mu}_1+S_1 K_2^{\mu}}{\gamma^2-S_1S_2},
\;\;\;\;\;K_2^{\flat,\mu}=-\gamma \alpha'\frac{\gamma K_2^{\mu}+S_2 K^{\mu}_1}{\gamma^2-S_1S_2},&
\nonumber\\
&\gamma_{\pm}=-(K_1\cdot K_2)\pm\sqrt{\Delta},\;\;\;\;\;\;\Delta=-\det(K_i\cdot K_j)=(K_1 \cdot K_2)^2-K_1^2K_2^2,&
\nonumber\\
&\alpha=\frac{S_2\left(S_1-\gamma \right)}{\left(S_1 S_2-\gamma^2\right)},\;\;\;\;\;\;
\alpha'=\frac{S_1\left(S_2-\gamma\right)}{\left(S_1S_2-\gamma^2\right)} \, ,&\label{eq:def_gamma_3_mass}
\end{eqnarray}
with $S_i = K_i^2$.

By definition, any box terms containing our chosen triple cut will
also contain an additional propagator, $1/(l-K_4)^2$. Inserting
the parametrization for $l^{\mu}$ of
eq.~(\ref{eq:sol_for_l_in_3_mass}) into this additional propagator
leads to the development of two poles, $t_{\pm}$, in the
propagator $(l-K_4)^2\propto (1/t)(t-t_-)(t-t_+)$. We can take
advantage of the occurrence of these poles to separate the box
terms from the triangle pieces.

The numerators of these box poles are given by an effective
quadruple cut generated by the extra pole along with the original
triple-cut. As we have seen above, the quadruple cut of a box
actually corresponds to one of the two contributions in the
construction of a box coefficient. Therefore each box term can be
written as a sum of two residue terms,
$\sum_{i=\pm}d_{i}/(\chi_i(t-t_{i}))$, with $d_{i}$ the residue
corresponding to the box coefficient of the pole $t_{i}$ and
$\chi_i$ a constant factor depending upon the box in question.

Analytically, in order to remove these box terms we simply expand the parametrized triple cut integrand expression around $t\rightarrow \infty$. The box terms behave as $1/t\rightarrow0$ in this limit and thus drop out. Taking a parameter to infinity numerically is problematic, so instead we use a different approach in the computer code \texttt{BlackHat}~\cite{Berger:2008sj}. Considering $t$ as a complex parameter, the box terms appear as poles in the complex plane of $t$, whereas the triangle coefficient is at the origin of the $t$ plane. To remove the box terms we simply systematically ``clean'' the complex plane by subtracting all box pole terms from our triple cut expression. A final complication to both numerical and analytic approaches is the presence of the $1/t$ factor in the box propagator. To account for this we add back the sum of all box terms evaluated at $t=0$ , i.e. add $(d_--d_+)/(\chi_i(t_--t_+))$ for each box term. An alternative approach to this last step and further discussion on the analytic properties of the three-mass triangle can be found in~\cite{Forde:2007mi} and~\cite{BjerrumBohr:2007vu}. In addition, the application of on-shell recursion to the computation of certain triangle coefficients can be found in~\cite{Bern:2006dp}.

\begin{figure}[hp]
\centerline{\epsfig{file=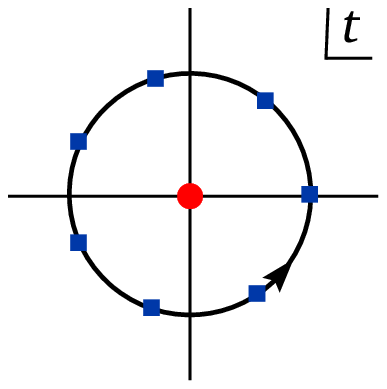,height=4cm}}
\caption{The points on the circle used by the discrete Fourier projection, cf. eq.~(\ref{eq:DFP}).}
\label{DFP}
\end{figure}

After the elimination of box terms we are left with a finite power series in $t$,
\begin{eqnarray}
&&\hspace*{-0.8cm}C(t) \equiv A_1(t)A_2(t)A_3(t) - \sum\limits_{i = \pm}
\frac{d_i}{\chi_i (t - t_i)} = \sum\limits_{j = -n}^n c_j \int dt \, t^j \, ,
\label{eq:tri_int_exp}
\end{eqnarray}
where the $c_j$ are rational coefficients.
The upper and lower limits $\pm n$ of the sum is determined by the theory in question, for example $n=3$ for renormalizable theories. To relate this sum of terms to the scalar triangle integral we first note that, for this particular parametrization, the integrals over any power of $t$ in eq.~(\ref{eq:tri_int_exp}) vanish as proved in~\cite{Forde:2007mi}. Then the  only remaining term, $c_0\int dt $, is in the form of a rational coefficient multiplying a scalar triangle integral. In an analytic formalism the series expansion around $t\rightarrow \infty$ will automatically isolate this term. So we can directly relate this sole remaining term to the desired coefficient of the basis triangle integral we have isolated with our triple cut. Numerically, our final step is to note that, since the power series in $t$ terminates at a finite power, $n$, the full contour integral is equivalent to a discrete Fourier projection~\cite{Berger:2008sj} with $2n+1$ evaluation points. The triangle coefficient is therefore given by,
\begin{eqnarray}
c_0=\frac{1}{2n+1}\sum_{j=-n}^{n}C \left(t_0e^{2\pi i\;j/(2n+1)}\right) \, .
\label{eq:DFP}
\end{eqnarray}
The arbitrary complex number $t_0$ is the radius of the numerical
Fourier projection, as illustrated in Fig.~\ref{DFP}. For
technical details we refer to Ref.~\cite{Berger:2008sj}.

\subsubsection{Bubbles}

The computation of bubble coefficients proceeds along similar
lines as above. A two-particle cut isolates a single bubble
coefficient, but will also capture triangles and boxes which share
the same cut. Again we use the differing analytic properties of
the terms with additional propagators to separate the triangle and
box terms from the bubble coefficient.

As before, we start from a specially chosen parametrization of the loop momentum. A two-particle cut leaves two free parameters,  $y$ and $z$. We then choose to parametrize the two-particle cut, bubble, loop-momenta as,
\begin{eqnarray}
&&\hspace*{-0.8cm}l_i^{\mu}(y,z) =
 \frac{1}{2}\, K_i^\mu
 + (y-\frac{1}{2}) \left(\Kf1^\mu-\chi^\mu\right)
 + \frac{z}{2}\,\Bmp{\Kfm1|\gamma^\mu|\chi^-}
 + \frac{y(1-y)}{2\,z}\langle\chi^-|\gamma^\mu|\Kfm1 \rangle\,.
 \nonumber\\
&&\label{TwoParticleParametrization}
\end{eqnarray}
Here $\chi$, an arbitrary massless momentum, is used to define the massless momentum $\Kf1^\mu = K_1^\mu - \chi^\mu$, with its normalization chosen so that $K_1\cdot \chi = K_1^2/2$.

The triangle and box terms sharing the two-particle cut will contain at least one additional propagator $1/(l-K_2)^2$. The parametrization of eq.~(\ref{TwoParticleParametrization}) will then introduce poles in $y$ or $z$. The residue of the poles in terms of $y$ is given by, effectively, a triple cut expression, which is a combination of the pole and the original two-particle cut. Contained inside this expression are a single triangle, isolated by the effective triple cut, and possibly box terms with the same cut.

Unfortunately, the more complicated structure of the momentum parametrization and the bubble integrand means that a simple extension of the triangle procedure is not so straightforward. This is because for the parametrization~(\ref{TwoParticleParametrization}) the integrals over positive powers of $y$ are non-zero, and given by,
\begin{eqnarray}
\int dy y^n=\frac{1}{n+1}\int dy\, .\label{eq:y_int_result}
\end{eqnarray}
The integrals over powers of $z$ still vanish. We must therefore
alter our approach for extracting the bubble coefficient. As
before, we discard any terms in the expansion around $z\rightarrow
\infty$  that depend upon $z$, but we must retain the coefficients
of all powers of $y$ (which are all guaranteed to be positive due
to our parametrization choice). For a renormalizable theory the
maximum power is $y^2$. We can relate these terms to the scalar
integral using eq.~(\ref{eq:y_int_result}), which is independent
of $y$ and $z$, $\int dy dz$. The sum of the resulting terms forms
{\it only} part of the bubble coefficient.

The source of the remaining contribution is the triangle expression from the residues of the poles in $z$.
This effective triple cut does not vanish in the double expansion in $y$ and $z$. Therefore, we obtain an additional contribution which needs to be subtracted from the part of the bubble coefficient computed in the previous paragraph.
Further details on this subtraction term can be found in Ref.~\cite{Forde:2007mi}.

To numerically extract the bubble coefficient we must first ``clean'' the complex plane of all pole terms. Computing the residues of each pole involves simply computing the triple cut at the location of each pole.
Once all pole terms have been removed we are free to extract the bubble coefficient from the remaining non-pole terms using a double discrete Fourier projection, in both $y$ and $z$. Instead of naively evaluating at as many points as there are coefficients in $y$ and $z$ we use the nature of our parametrization to reduce the number of points at which we need to evaluate $z$ by one.  The coefficient in a renormalizable theory is given by
\begin{equation}
b_0=\frac{1}{20}\sum_{j=0}^{4}\Bigg[B \left(y=0,z=t_0e^{2\pi i j/5}\right) + 3 B\left(y=2/3,z=t_0e^{2\pi i j/5}\right)\Bigg] \, .
\end{equation}
$B(y,z)$ denotes the two-particle cut from which triangles and boxes have been subtracted that share this cut.
The more general case is given in~\cite{Berger:2008sj}. Alternative approaches to the computation of the bubble coefficients have been proposed in the literature, such as Mastrolia's use of Stokes Theorem~\cite{Mastrolia:2009rk}.

\subsubsection{Massive particles}

The expressions we have given above are for amplitudes with purely massless particles. The addition of
massive particles which do not circulate in the loop is straightforwardly accommodated within the above,
with no changes. Including massive particles inside the circulating loop requires further exposition.
Two changes are required, firstly the loop momentum parametrization needs to be extended to include
massive particles. Secondly we must also compute the coefficients of tadpole integrals in addition to
the bubbles, triangles and boxes. A detailed discussion of the extension of the procedure to the computation
of massive particles as well as the computation of the tadpole terms themselves is given in~\cite{Kilgore:2007qr}.
In addition there is the problem of the wave-function renormalization with massive particles, which has been addressed in Ref.~\cite{Ellis:2008ir}.

\subsection{Extraction of Integral Coefficients at the Integrand Level}
\label{section:OPP}

So far we have described methods which used analytic limits or a
combination of the  subtraction of poles on the complex plane and
discrete Fourier projections. An alternative approach developed by
Ossola, Papadopoulos and Pittau (OPP) uses the knowledge of the
general structure of the integrand $\tilde{A}(l)$ instead. The
integrand is built up from a standard set of terms. These terms
either vanish upon integration or correspond to one of the scalar
integral basis functions of eq.~(\ref{integralbasis}). Computing
the scalar basis integral coefficients then reduces to the problem
of finding the coefficients of the OPP decomposition of the
integrand.

In four dimensions, the integrand can be written as,
\begin{eqnarray}
\tilde{A}_n(l)&=&\sum_{1\leq i_1< i_2<i_3<i_4\leq n}\frac{d_{i_1i_2i_3i_4}(l)}{D_{i_1}D_{i_2}D_{i_3}D_{i_4}}
+\sum_{1\leq i_1< i_2<i_3\leq n}\frac{c_{i_1i_2i_3}(l)}{D_{i_1}D_{i_2}D_{i_3}}
\nonumber\\
&&+\sum_{1\leq i_1< i_2\leq n}\frac{b_{i_1i_2}(l)}{D_{i_1}D_{i_2}}+\sum_{1\leq i_1\leq n}\frac{a_{i_1}(l)}{D_{i_1}},
\label{decomposition}
\end{eqnarray}
with the propagator $D_i=(l-K_i)^2-m_i^2$, where the mass of the
cut propagator with momentum $l_i$ has mass $m_i$.  From now on we include
massive particles in the discussion. The form of the numerators in
eq.~(\ref{decomposition}) depend upon the basis we choose for the
loop momenta. We wish to choose this momentum basis so that each
scalar integral basis coefficient of $B_4$ corresponds to a single
term in the integrand decomposition and the integrals over the
remaining structures vanish. In the form originally presented by
OPP~\cite{Ossola:2006us}, a momentum parametrization for the box,
triangle and bubble very similar to eqs.~(\ref{eq:box_solutions}),
(\ref{eq:sol_for_l_in_3_mass}) and
(\ref{TwoParticleParametrization}) was used.

An alternative momentum parametrization, presented by Ellis et
al.~\cite{Ellis:2007br},  is related to the van Neerven-Vermaseren
basis~\cite{vanNeerven:1983vr}. The generic form of a momentum in
this basis is
\begin{eqnarray}
l^{\mu}=\sum_{j=1}^{D_P}v^{\mu}_{j}+\sum_{j=1}^{D_T} \alpha_j n^{\mu}_{j}.\label{eq:GKM_mom_param}
\end{eqnarray}
This is a decomposition into to two sets of basis vectors. The
vectors $v^{\mu}_{j}$ span the physical space defined by the
external legs $K_i$ and the vectors $n^{\mu}_{j}$ span the space
transverse to this physical space. For a box $D_T=1$ and $D_P=3$,
for a triangle $D_T=2$ and $D_P=2$ and for a bubble $D_T=3$ and
$D_P=1$. The basis vectors are chosen such that $n_{i}\cdot
n_{j}=\delta_{ij}$, $n_{i}\cdot K_j=0$ and $n_{i}\cdot v_j=0$. The
$v_j$'s are chosen such that any cut legs are on-shell.

The numerators of the propagator terms are then arranged in the
following way~\cite{Ellis:2007br},
\begin{eqnarray}
d_{i_1i_2i_3i_4}(l)&=&c^{0}_{i_1i_2i_3i_4}+c^{1}_{i_1i_2i_3i_4}t_{1}, \label{eq:OPP_coeff_breakdown}\\
c_{i_1i_2i_3}(l)&=&c^{0}_{i_1i_2i_3}
+c^{1}_{i_1i_2i_3}t_{1}+c^{2}_{i_1i_2i_3}t_{2}+c^{3}_{i_1i_2i_3}(t_{1}^2-t_{2}^2)
\label{eq:OPP_coeff_breakdowntri} \\
&&+t_{1}t_{2}(c^{4}_{i_1i_2i_3}+c^{5}_{i_1i_2i_3}t_{1}+c^{6}_{i_1i_2i_3}t_{2}),\nonumber\\
 b_{i_1i_2}(l)&=&c^{0}_{i_1i_2}+c^{1}_{i_1i_2}t_{1}+c^{2}_{i_1i_2}t_{2}+c^{3}_{i_1i_2}t_{3}
 +c^{4}_{i_1i_2}(t_{1}^2-t_{3}^2)
 \label{eq:OPP_coeff_breakdownbub}\\
 &&+c^{5}_{i_1i_2}(t_{2}^2-t_{3}^2)+c^{6}_{i_1i_2}t_{1}t_{2}
 +c^{7}_{i_1i_2}t_{1}t_{3}+c^{8}_{i_1i_2}t_{2}t_{3}.\nonumber \\
 a_{i_1}(l)&=& c^{0}_{i_1}+c^{1}_{i_1}t_{1}+c^{2}_{i_1}t_{2}+c^{3}_{i_1}t_{3}+c^{4}_{i_1}t_{4} \, ,
\label{eq:OPP_coeff_breakdowntad}
\end{eqnarray}
for the tadpole, bubble, triangle and box, $i=1,2,3,4$
respectively. Here the $t_{j}=(n_{j}\cdot l)$ depend on the specific parametrization
of the loop momentum, which is chosen differently for boxes, triangles, bubbles and tadpoles.
 The coefficients we
wish to compute are the $c^{0}$ terms which correspond to the
scalar basis integral coefficients. We will not discuss the
computation of the tadpole terms here, methods to compute these
can be found
in~\cite{Ossola:2006us,Britto:2009wz,Kilgore:2007qr,Ellis:2008ir,Ellis:2007br}.
The remaining coefficients multiply $t_{j}$ terms (or combinations
of such terms), which vanish upon integration over $l$. As in
section~\ref{section:genuni}, the choice of the representation for
the loop momentum is crucial in this approach.

Trying to solve for the entire set of coefficients at once
is clearly not the optimal approach. Using unitarity cuts we can
isolate individual terms of the integrand. The application of all
possible cuts allows us to systematically solve for all of the
coefficients sequentially. Isolating a single box term $d_{i_1i_2i_3i_4}$ with
a quadruple cut works in the same way as in
section~\ref{section:genuni}. Again we have two solutions for the
completely frozen box loop momentum. For the momentum basis
corresponding to the integrand decomposition
eq.~(\ref{eq:OPP_coeff_breakdown}) these are
\begin{eqnarray}
l^{\mu}_{\pm}&=&v^{\mu}_{1}+v^{\mu}_{2}+v^{\mu}_{3}\pm\sqrt{m_0^2-(v_{1}+v_{2}+v_{3})^2}\,n^{\mu}_{1}.
\end{eqnarray}
As above, the box coefficients in the OPP basis are given by
products of four tree amplitudes,
\begin{eqnarray}
c^{0}_{i_1i_2i_3i_4}\!\!\!&=&\!\!\! \frac{1}{2} \sum\limits_{a = \pm} d_a, \qquad
d_a = A_1(l_{a})A_2(l_{a})A_3(l_{a})A_4(l_{a})\, ,
\\
c^{1}_{i_1i_2i_3i_4}\!\!\!&=&\!\!\! \frac{1}{2} \left( d_{+} - d_{-} \right) \, .
\nonumber
\end{eqnarray}
Here, $c^{0}_{i_1i_2i_3i_4}$ is the desired scalar box integral
coefficient, $d_0$ (cf. eq.~(\ref{eq:box_coeffs})). The other
coefficient is needed for the computation of the triangle
coefficients.

For the computation of the triangle coefficients we apply a triple
cut to isolate the particular triangle we are interested in. Much
like before this means that we also have box terms polluting the
result. To solve this problem we subtract the complete box
contribution from the integrand. So in order to find the triangle
 coefficients $c^k_{i_1 i_2 i_3}$ we do not evaluate
$A_1(l)A_2(l)A_3(l)$, but instead,
\begin{eqnarray}
A_1(l)A_2(l)A_3(l)-\sum_{i_4}\frac{d_{i_1i_2i_3i_4}(l)}{D_{i_4}},
\end{eqnarray}
at different choices for the loop momentum of the triangle. This
is in contrast to the procedure of the previous section. There we
subtracted the residue of the box pole from the triangle rather
than the entire contribution of the box (\ref{eq:box_coeffs}) from
the integrand as above. The triangle momentum parametrization
corresponding to the integrand structure
eq.~(\ref{eq:OPP_coeff_breakdowntri}) contains one free parameter
which we label $\alpha$,
\begin{eqnarray}
l^{\mu}=v^{\mu}_{1}+v^{\mu}_{2}+\sqrt{m_0^2-\alpha^2-(v_{1}+v_{2})^2}\,n^{\mu}_{1}+\alpha \,n^{\mu}_{2}.
\end{eqnarray}
There are seven unknown coefficients in
eq.~(\ref{eq:OPP_coeff_breakdowntri}),  so we need to evaluate
this at seven, in principle arbitrary, values of $\alpha$ to
determine all coefficients. For increased numerical stability,
however, they can be chosen to lie on a circle as in the previous
section. The resulting set of linear equations can then be solved
to find the full set of coefficients, which is needed in the
computation of the bubble coefficient.

Finally, in order to compute the bubbles we apply a double cut,
which again isolates a single bubble but also any triangle and box
coefficients that share the same cut. Again we simply subtract the
unwanted terms from the integrand to remove them. To find the
coefficients $b^k_{i_1i_2}$ we evaluate
\begin{eqnarray}
A_1(l)A_2(l)-\sum_{i_3}\frac{c_{i_1i_2i_3}(l)}{D_{i_3}}-
\frac{1}{2!}\sum_{i_3i_4}\frac{d_{i_1i_2i_3i_4}(l)}{D_{i_3}D_{i_4}},
\end{eqnarray}
at different values of the loop momentum of the bubble. Here the
bubble loop  momentum corresponding to the  parametrization
eq.~(\ref{eq:OPP_coeff_breakdownbub}) is,
\begin{eqnarray}
l^{\mu}=v^{\mu}_{1}+\sqrt{m_0^2-\alpha_1^2-
\alpha_2^2-v_{1}^2}\,n^{\mu}_{1}+\alpha_1\, n^{\mu}_{2}+\alpha_2 \,n^{\mu}_{3},
\end{eqnarray}
with two free parameters, $\alpha_1$ and $\alpha_2$. Solving for
the nine coefficients in eq.~(\ref{eq:OPP_coeff_breakdownbub})
requires nine linear equations. These can be generated by choosing
nine different values of $l$ via the choice of the two free
parameters. The full set of nine coefficients is only required if
we wish to compute the tadpole coefficients, which only appear if
massive particles are present in the loop. Further details can be
found in Refs.~\cite{Ellis:2008ir,Ellis:2007br}.

\subsection{Rational Terms from D-Dimensional Generalized Unitarity}
\label{section:D-Dimensional_Generalized_Unitarity}

We have so far restricted ourselves to working with cut legs in
four dimensions, keeping the rational terms out of reach.
Considering cuts in $D$ dimensions allows us to use the $B_D$
integral basis of eq.~(\ref{integralbasis}), bringing the
computation of all terms within our grasp. This is related to van Neerven's important observation that dispersion relations for Feynman integrals converge in dimensional regularization~\cite{vanNeerven:1985xr}. There are multiple
different approaches to go beyond four-dimensional cuts as already
proposed in Refs.~\cite{Bern:1995db}. The additional $D-4$
components can be related to massive terms in four
dimensions~\cite{Bern:1995db}. In a related approach the $D-4$
components can be converted to an additional integral which can be
used to compute the full amplitude~\cite{Anastasiou:2006gt}.

Alternatively, we can directly extend the approaches of either
Section~\ref{section:genuni} or Section~\ref{section:OPP}, as we
will now describe, starting with the generalization of
Section~\ref{section:OPP}.

\subsubsection{D-Dimensional Unitarity at the Integrand Level}

OPP have suggested a two-step computational procedure. Here, the
$D-4$ terms in the numerator of the integrand are computed using a
separate set of Feynman diagrams. The corresponding Feynman rules
have been derived for QCD and electroweak theories in a series of
papers~\cite{OPPrat,Garzelli:2009is}. The $D-4$ contributions from
the denominator of the integrand are found by computing
coefficients of an extended OPP basis structure for the
integrand~\cite{Ossola:2006us,OPPrat}.

A direct extension of  the OPP approach to $D$ dimensions was
derived by Giele et al.~\cite{Giele:2008ve}. This approach
combines trees in higher dimensions with an extended integrand
basis for the one-loop integrand, taking into account the
additional structure in higher dimensions. Again as in
Section~\ref{section:OPP} the choice of the loop momentum
determines the form of the integrand structures. The form of the
higher dimensional loop momenta can be decomposed into a
four-dimensional part $\overline{l}$ and an orthogonal
$(D-4)$-dimensional part $\tilde{l}$,
$l^{\mu}=\overline{l}^{\mu}+\tilde{l}^{\mu}$. The on-shell
constraint then means that $l^2=\overline{l}^2+\tilde{l}^2$ and so
the four-dimensional component effectively becomes massive with
mass $-\tilde{l}^2=\mu^2$, where $\mu^2$ is the ``scale'' of the
higher-dimensional subspace.

Since the external momenta remain in four dimensions, we find that
there are a limited number of additional integrand structures that
can be present because the external momenta are orthogonal to the
$D-4$ additional transverse dimensions. The numerator structures
can only depend upon the higher-dimensional terms through even
powers of $\mu$. Similarly, there can be at most one additional
cut leg, since $\mu^2$ is fixed by the fifth constraint. Therefore
we must include a pentagon in the integrand basis and so for a
renormalizable theory the extended integrand structures are given
by,
\begin{eqnarray}
\tilde{A}_n(l)&=&\sum_{1\leq i_1< i_2<i_3<i_4<i_5\leq n}\frac{\tilde{e}_{i_1i_2i_3i_4i_5}(l)}{D_{i_1}D_{i_2}D_{i_3}D_{i_4}D_{i_5}}
+\sum_{1\leq i_1< i_2<i_3<i_4\leq n}\frac{\tilde{d}_{i_1i_2i_3i_4}(l)}{D_{i_1}D_{i_2}D_{i_3}D_{i_4}}
\nonumber\\
&&+\sum_{1\leq i_1< i_2<i_3\leq n}\frac{\tilde{c}_{i_1i_2i_3}(l)}{D_{i_1}D_{i_2}D_{i_3}}+
\sum_{1\leq i_1< i_2\leq n}\frac{\tilde{b}_{i_1i_2}(l)}{D_{i_1}D_{i_2}}+\sum_{1\leq i_1\leq n}\frac{\tilde{a}_{i_1}(l)}{D_{i_1}}.
\end{eqnarray}
Here the numerator coefficients are given by,
\begin{eqnarray}
\tilde{e}_{i_1i_2i_3i_4i_5}(l)&=&c^{0}_{i_1i_2i_3i_4i_5},\label{eq:GKM_coeff_breakdown}\\
\tilde{d}_{i_1i_2i_3i_4}(l)&=&d_{i_1i_2i_3i_4}(\overline{l})+\mu^2(c^{2}_{i_1i_2i_3i_4}+t_{1}\,c^{3}_{i_1i_2i_3i_4})+\mu^4\,c^{4}_{i_1i_2i_3i_4}, \nonumber\\
\tilde{c}_{i_1i_2i_3}(l)&=&c_{i_1i_2i_3}(\overline{l})+\mu^2\left(t_{1}\, c^{8}_{i_1i_2i_3}+t_{2}\,c^{9}_{i_1i_2i_3}+c^{10}_{i_1i_2i_3}\right),
\nonumber\\
 \tilde{b}_{i_1i_2}(l)&=&b_{i_1i_2}(\overline{l})+\mu^2 c^{10}_{i_1i_2},
 \nonumber\\
 \tilde{a}_{i_1}(l)&=&a_{i_1}(\overline{l}) \, .\nonumber
\end{eqnarray}
where again $t_j = n_j\cdot l$.

The attentive reader may be worried about how this basis can be
used if we have a fractional number of additional dimensions, for example
when $D=4-2\varepsilon$. This problem can be sidestepped by noting
that the dependence of the amplitudes upon the $(D-4)$-dimensional
subspace is {\it linear}. So the fractional dimensional structure
of the amplitude can be reconstructed by evaluating the amplitude
at two different integer dimensions. Using $A^{\rm
1-loop}_D=A^{\rm 1-loop}_{0}+(D-4)A^{\rm 1-loop}_{1}$ the full
$D$-dimensional amplitude can then be reconstructed~\cite{Giele:2008ve}. This allows
us to freely choose the exact form of the regularization scheme.
The integer dimension chosen when evaluating the trees that enter
the calculations is arbitrary up to the constraint that the
dimension has to be an even number if fermions are present in the loop.

The final step is to relate the new integrand structures of
eq.~(\ref{eq:GKM_coeff_breakdown}), proportional to powers of
$\mu$, to their contribution in eq.~(\ref{integralbasis}). As
before, terms proportional to $t_{j}$ terms will vanish and so we
are left with terms proportional only to $\mu$. Scalar integrals
multiplied by powers of $\mu$ can be related to scalar integrals
in higher dimensions. From eq.~(\ref{eq:GKM_coeff_breakdown}) we
have the following different integrals,
\begin{eqnarray}
\int d^Dl\frac{\mu^2}{D_{i_1}D_{i_2}D_{i_3}D_{i_4}}&=&-\frac{D-4}{2}I^{D+2}_{i_1i_2i_3i_4}\stackrel{D\rightarrow4}{\rightarrow}0
\nonumber\\
\int d^Dl\frac{\mu^4}{D_{i_1}D_{i_2}D_{i_3}D_{i_4}}&=&\frac{(D-4)(D-2)}{4}I^{D+4}_{i_1i_2i_3i_4}\stackrel{D\rightarrow4}{\rightarrow}-\frac{1}{6}
\nonumber\\
\int d^Dl\frac{\mu^2}{D_{i_1}D_{i_2}D_{i_3}}&=&-\frac{(D-4)}{2}I^{D+2}_{i_1i_2i_3}\stackrel{D\rightarrow4}{\rightarrow}-\frac{1}{2}
\nonumber\\
\int d^Dl\frac{\mu^2}{D_{i_1}D_{i_2}}&=&-\frac{(D-4)}{2}I^{D+2}_{i_1i_2}\stackrel{D\rightarrow4}{\rightarrow}\frac{m^2_{i_1}+m^2_{i_2}}{2}-\frac{1}{6}S_{i_1 i_2}. \label{eq:Ddimrational}
\end{eqnarray}
where $S_{i_1i_2}$ is the mass of the bubble labelled by $i_1$ and
$i_2$. Examining
the limits of these integrals as we return to four dimensions, we
see that the integrals are finite and correspond to purely
rational numbers. Each new integrand factor therefore contributes
to the rational term.

Computing the one-loop amplitude is  now very similar to the
cut-computation procedure of Section~\ref{section:OPP}. Starting
with the pentagon we find that the loop momenta will be completely
frozen by the four cuts on the four-dimensional component of
$l_5^{\mu}$ and the constraint on $\mu^2 = -\tilde{l}^2$ from the
fifth propagator $(\overline{l}+\tilde{l}-K_5)^2=m_5^2$. The
resulting expression is then inserted into the penta-cut so that
we find the pentagon coefficient using,
\begin{eqnarray}
c^{0}_{i_1i_2i_3i_4i_5}=A_1(l)A_2(l)A_3(l)A_4(l)
A_5(l).
\end{eqnarray}
The trees must be evaluated, as described above, at two different
integer dimension choices for the internal loop momentum. The
pentagon  coefficient is proportional to $D-4$ and so will vanish
in the four-dimensional limit. We need it only for subtraction
when computing the box terms as we will now explain.

The box is computed as before but with the pentagon subtracted,
\begin{eqnarray}
A_1(l)A_2(l)A_3(l)A_4(l)-\sum_{i_5}\frac{e_{i_1i_2i_3i_4i_5}(l)}{D_{i_5}}.
\end{eqnarray}
With three additional coefficients to be determined, compared to
the four-dimen\-sional case, we need three additional evaluation
points in order to generate enough equations to solve for all
coefficients. All coefficients are required for the computation of
the triangle terms. Computing the triangle and bubble
contributions to the rational terms follows in a similar vein as
for the boxes and we direct the reader to Ref.~\cite{Giele:2008ve}
for further details on the computation of all such contributions.

\subsubsection{Four-dimensional Generalized Unitarity with a
$D-4$-Dimen\-sional Mass}

The differing dependence of the rational terms of
eq.~(\ref{eq:GKM_coeff_breakdown}) on $\mu^2$ suggests that we
could use the analytic structure of the integrand to compute the
rational terms at the integral level. This approach of
Badger~\cite{Badger:2008cm} is similar in spirit to the extraction
of the cut-coefficients described in Section~\ref{section:genuni}.
Here the loop momentum is massive and four-dimensional rather than
massless and $D$-dimensional.

To compute the rational contribution of the box, we start from the
quadruple cut expression of a box. Now each cut leg has
an additional mass $\mu^2$. This cut isolates a single box
coefficient and also any pentagon terms which share the same cut.
Schematically we have,
\begin{eqnarray}
A_1(l(\mu^2))A_2(l(\mu^2))A_3(l(\mu^2))A_4(l(\mu^2))=r_1+r_2\mu^2+r_3\mu^4+\sum_{i}\frac{e^0_i}{\chi_i(\mu^2-\mu^2_i)}.\label{eq:rat_box_comp}
\end{eqnarray}
where $\mu^2_i$ is the pole in $\mu^2$ for the pentagon $i$,
$\chi_i$ is a constant factor depending on the pentagon in question.
The additional propagator of the
pentagon terms shows up as a $1/(\mu^2-\mu^2_i)$ factor. We wish
to separate these terms as well as the other terms in the powers
series in $\mu^2$ from each other. Here we only want the
coefficient $r_3$ of $\mu^4$, which is the term that contributes
to the purely rational part, cf. eq.~(\ref{eq:Ddimrational}). So
simply expanding
$A_1(l(\mu^2))A_2(l(\mu^2))A_3(l(\mu^2))A_4(l(\mu^2))/\mu^4$ around the
limit $\mu^2 \rightarrow \infty$ will
give us the coefficient directly.

Moving on, performing a triple cut isolates a single triangle as
well as boxes and pentagons. These objects contain poles in both
$\mu^2$ and the unconstrained component of the loop momenta, $t$.
We are only interested in the coefficient of $\mu^2$ and so series
expanding around both $t\rightarrow \infty$ and $\mu^2\rightarrow
\infty$ isolates this single term. For the bubbles a similar
procedure applies now in three parameters, a complete description
is given in~\cite{Badger:2008cm}.

As in Section~\ref{section:genuni} taking an infinite limit
numerically is difficult to do. Similar to the numerical approach
to the computation of the cut-containing terms described in
Sect.~\ref{section:genuni}, we can adapt Badger's method. This is
the approach implemented within
\texttt{BlackHat}~\cite{Berger:2008sj}. Starting from
eq.~(\ref{eq:rat_box_comp}) we ``clean'' the complex $\mu$ plane
by subtracting all pentagons at their poles. We then perform a
discrete Fourier projection in $\mu$ to extract the rational
contribution $r_3$. For the rational contribution of a triangle we
consider the triple cut expression and then clean the complex
$\mu$ plane by subtracting all terms with an additional
propagator. Each such term has a pole in $t$ and its numerator is
a quadruple cut. Here we are subtracting the quadruple cut residue
of the additional propagator and not the integrand box and
pentagon terms, as we would in the OPP approaches. We sidestep
therefore any loss of numerical stability that would arise from
cancellations between these box and pentagon terms since these
pieces are never separated out in the quadruple cut. The
computation of the rational contribution from the two-particle cut
of a bubble follows along a similar line.

\section{On-Shell Recursion at One Loop}
\label{recursion}

As above, an alternative approach to obtain the full amplitude is
to use four-dimensional unitarity methods for the
cut-constructible terms and to use recursion relations for the
purely rational remainder. The recursive approach developed in
Refs.~\cite{Bern:2005cq,Berger:2006ci} has already been reviewed
extensively in \cite{Bern:2007dw}. While this approach is very
useful for analytical calculations and even allows to obtain
closed-form all-multiplicity results~\cite{Berger:2006vq}, it is
not straightforward to cast into a form suitable for numerical
implementation. This is due to the removal of spurious
singularities via the introduction of overlap terms, which is
difficult to perform in an automated fashion. The approach has
subsequently been modified to allow efficient numerical
implementation into the automated C++ library
\texttt{BlackHat}~\cite{Berger:2008sj}. We will discuss here this
modified approach, for amplitudes with massless particles in the
loop.

We begin by dividing the amplitude into cut-constructible and rational terms, as in
eq.~(\ref{integralbasis}). The rational terms are defined by setting all
scalar integrals to zero,
\begin{eqnarray}
A_n^{\oneloop} & = & \sum\limits_{j \in B_4} c^{D=4}_j \mathcal{I}^D_j
+ {\cal R}_n \, , \\
C_n & \equiv & \sum\limits_{j \in B_4} c^{D=4}_j \mathcal{I}^D_j, \quad
\mathcal{R}_n \equiv \left. A_n^{\oneloop} \right|_{\mathcal{I}^D_j \rightarrow 0} \, .
\label{rational}
\end{eqnarray}
In the following we assume that the cut-containing terms have been computed via the
generalized unitarity methods described in the previous section.

The basic idea is to complex continue $\mathcal{R}$ and use Cauchy's theorem to
reconstruct the rational term from its poles in the complex plane, similarly
to the tree-level approach introduced in Section~\ref{treerecursion}. We add this rational
term
to the previously-computed cut terms $C$ to obtain the full physical
amplitude at $z = 0$.
However, two
of the basic premises of the derivation in Section~\ref{treerecursion}
do not hold in general at one loop:
For one, we cannot always
find shifts~(\ref{SpinorShift}) such that the amplitude vanishes as $z \rightarrow \infty$.
And for another, the division of the amplitude into cut and rational
parts introduces the presence of spurious, unphysical poles in the complex plane when considering
the rational part separately.
That is, the rational part has the following form upon complex continuation,
\begin{equation}
\mathcal{R}_n(z) = \mathcal{R}_n^P(z) + \mathcal{R}_n^S(z) + \mathcal{R}_n^{\mbox{\tiny large\,}\tiny z} \, ,
\end{equation}
where $\mathcal{R}_n^P$ denotes the contribution from physical poles, which are as at the tree level
simple poles, whereas the spurious poles can be simple or double poles, and the contributions
$\mathcal{R}_n^{\mbox{\tiny large\,} z}$ display polynomial growth in $z$:
\begin{eqnarray}
\mathcal{R}_n^P(z) & = & \sum\limits_\alpha \frac{A_\alpha}{z-z_\alpha} \, , \\
\mathcal{R}_n^S(z) & = & \sum\limits_\beta \left( \frac{B_\beta}{(z-z_\beta)^2} +
 \frac{C_\beta}{z-z_\beta} \right) \, , \nonumber \\
\mathcal{R}_n^{\mbox{\tiny large\,} \tiny z} & = & \sum\limits_{\sigma = 0}^{\sigma_{\mbox{\tiny max}}}
D_\sigma z^\sigma \, . \nonumber
\end{eqnarray}
Here, $A, B, C,$ and $D$ are functions of the external momenta.
We postpone the case where $\mathcal{R}_n^P(z)$ has a more complicated structure
to Section~\ref{sec:infinity}.

\begin{figure}[hp]
\centerline{\epsfig{file=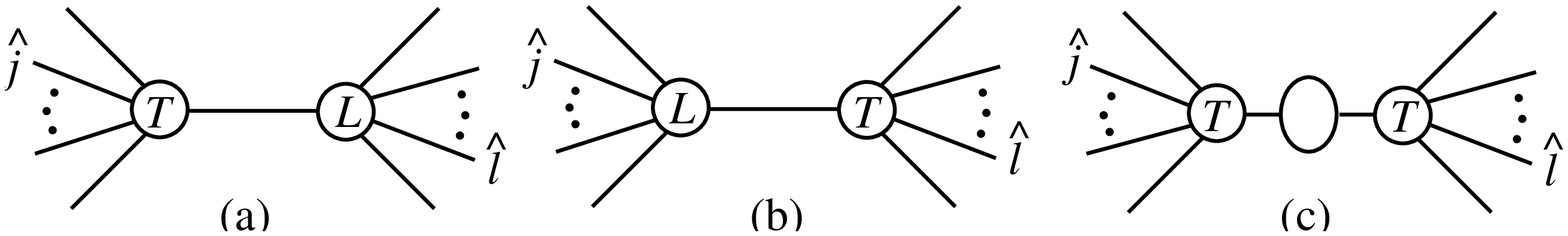,height=2.5cm}}
\caption{Schematic representation of one-loop recursive
contributions to eq.~(\ref{RationalRecursion}).
The labels `T' and `L' refer to tree and the rational part of loop vertices,
respectively.}
\label{onelooprecfig}
\end{figure}

The physical contribution is recursively computed as in the tree level case,
\begin{eqnarray}
 - \sum_{{\rm poles}\ \alpha} \Res_{z=z_\alpha} {\mathcal{R}^P_n(z)\over z}
 &=& \hskip -.1cm
 \sum_{r,s}\, \sum_{h} \Biggl\{
 A^{h}_L(z= z_{rs}) {i\over K_{r\ldots s}^2}
\mathcal{R}_R^{-h}(z = z_{rs})
  \label{RationalRecursion}   \\
&& \qquad \quad + \,
\mathcal{R}_L^h(z = z_{rs}) {i\over K_{r\ldots s}^2}
A^{ -h}_R(z= z_{rs})
\nonumber \\
&& \qquad \quad + \,
A^{h}_L(z= z_{rs}) {i \mathcal{F}(K_{r\ldots s}^2) \over K_{r\ldots s}^2}
A^{-h}_R(z= z_{rs}) \Biggr\}
 \,. \nonumber
\end{eqnarray}
Here, we have expressed the rational
term as a sum of products of rational terms from lower-point amplitudes (defined according
to eq.~(\ref{rational})) with lower-point tree amplitudes.
The last term with $\mathcal{F}$ corresponds to a one-loop correction to the
propagator. Eq.~(\ref{RationalRecursion}) is illustrated in Fig.~\ref{onelooprecfig}.

The full amplitude is found by combining all rational and cut contributions,
\begin{equation}
A_n^{\oneloop}(0) =  C_n(0)   -\sum\limits_{{\rm poles}\ \alpha}
\Res_{z=z_\alpha}  {{\cal R}_n^P(z)\over z} +  \mathcal{R}_n^S(0) + \mathcal{R}_n^{\mbox{\scriptsize large\,\,} \tiny z} \, . \label{totalamp}
\end{equation}
We will now show how to obtain the last two contributions, starting with the spurious pole contribution.

\subsection{Spurious Poles}

The division into cut and rational parts
introduces spurious singularities in each of these terms which however cancel in the
full amplitude. These spurious singularities are already present in real kinematics.
 Cauchy's theorem requires us to sum over all poles,
whether physical or unphysical. In Ref.~\cite{Berger:2006ci} this problem was remedied
by adding additional rational terms  to the cut part.
These rational terms are constructed such that the cut and the rational parts individually
do not contain spurious singularities upon continuation into the complex plane.
However, these extra rational terms then contribute additional terms to the residues
at the physical poles in eq.~(\ref{RationalRecursion}) which
 have to be
subtracted from the rational part $\cal R$ in the recursive
construction by so-called overlap terms in order to avoid double
counting~\cite{Berger:2006ci}. This approach leads to compact
expressions in analytical calculations but is not particularly
amenable to numerical implementation.

An alternative way of dealing with the spurious singularities is to make use of the fact that
we know that they cancel in the full amplitude. In other words, we can extract
the spurious residues from the known cut parts,
\begin{equation}
 \mathcal{R}_n^S(0) = - \sum\limits_{\mbox{\scriptsize spur poles}\, \beta}
\Res_{z=z_\beta}  \frac{\mathcal{R}_n^S(z)}{z} =
+ \sum\limits_{\mbox{\scriptsize spur poles}\, \beta}
\Res_{z=z_\beta}  \frac{C_n(z)}{z} \, ,
\end{equation}
where $C_n(z)$ is the complex continued cut part. The spurious poles in $C_n(z)$
come from the vanishing of complex continued Gram determinants, $\Delta(z)$,
associated with bubble, triangle, and box integrals. Since the spurious poles
cancel between the rational and the cut parts, the spurious contribution
to the residues from the cut part can only be rational. To compute the
residue we therefore series expand the logarithms and polylogarithms
around the location of the vanishing Gram determinants and obtain a
series of rational functions. The spurious contribution is thus given by
evaluating
\begin{equation}
\mathcal{R}_n^S(0) =  \, \sum\limits_{\Delta_m(z_\beta) = 0}
\Res_{z=z_\beta} \left[ \sum\limits_j \frac{c^{D=4}_j(z) \left.
\mathcal{I}^D_j(z)\right|_{\mbox{\scriptsize rat}}}{z} \right] \equiv
\, \sum\limits_{\Delta_m(z_\beta) = 0}
\Res_{z=z_\beta} \frac{E_n^\beta(z)}{z} \, , \label{spuriousfinal}
\end{equation}
where the subscript ``rat'' indicates that we take the rational part of
the series expansion of the integrals
around the spurious poles. We
have introduced the abbreviation $E_n^\beta$ for these rational terms.
 The spurious poles $z_\beta$ are located where the shifted Gram determinants vanish,
$\Delta_m(z_\beta) = 0$, with $m = 2,3,4$ for bubble, triangle, and box integrals, respectively.
Note that poles in Gram determinants of
box integrals will in general also appear in the daughter triangle and bubble
integrals\footnote{By daughter integrals we
mean integrals that can be found from the parent integral by collapsing
one or more of the loop propagators.}.
Singularities in triangle integrals will feed down into the daughter bubble
coefficients, but not affect the parent box integrals.

The expansion of an integral around the location where its Gram determinant vanishes can be
obtained~\cite{expansion} by using the ``dimension-shifting relation''~\cite{dimshift} iteratively,
\begin{equation}
\mathcal{I}^D_j = \frac{1}{2} \left( \sum\limits_{i = 1}^j c_i\, \mathcal{I}^D_{j-1} [i] + (j-1-D) c_0\,
\mathcal{I}_j^{D+2} \right) \, , \label{dimshifteq}
\end{equation}
where $\mathcal{I}^D_{j-1} [i]$ denotes the lower-point integral obtained from
$\mathcal{I}^D_j$ by removing the $i$th propagator. The coefficients $c_i$ and $c_0$ are given by
\begin{equation}
c_i = \sum\limits_{k = 1}^j \left( Y^{-1} \right)_{ik} \, , \quad c_0 = \sum\limits_{k = 1}^j c_i
\sim \Delta_j \, ,
\end{equation}
where $Y^{-1}$ denotes the inverse of the modified Cayley matrix. These modified Cayley
matrices are listed for example
in \cite{Ellis:2007qk}. As indicated, the coefficient $c_0$ is proportional to the Gram
determinant. We obtain the series expansion of $\mathcal{I}_j^D$ in terms of its
Gram determinant,
\begin{equation}
\left. \mathcal{I}_j^D \right|_{\mbox{\scriptsize rat}} = \sum\limits_{k = 0} r_k \Delta_j^k \, ,
\end{equation}
with rational coefficients $r_k$ that are found by using
eq.~(\ref{dimshifteq}) iteratively. The explicit expression for
the rational expansion of the three-mass triangle is listed in
Ref.~\cite{Berger:2008sj}. Similar expansions for the remaining
integrals can be obtained as described above and will be listed in
a forthcoming publication (C.F. Berger et al., in preparation).

Numerically, we can evaluate eq.~(\ref{spuriousfinal}) by using a discrete Fourier sum,
which here only approximates the contour integral.
We evaluate the quantity given in the square bracket of (\ref{spuriousfinal}), $E_n^\beta(z)/z$,
 at $m$ points equally spaced around a circle
of radius $\delta_\beta$ in the $z$ plane, centered on the pole location $z_\beta$,
\begin{equation}
\mathcal{R}_n^S(0) \approx \frac{1}{m} \sum\limits_\beta \sum\limits_{j = 1}^m
\delta_\beta e^{2 i \pi j/m} \frac{E^\beta_n(z_\beta + \delta_\beta e^{2 i \pi j/m})}{
z_\beta + \delta_\beta e^{2 i \pi j/m}} \, .
\end{equation}
The sum over $\beta$ runs over all locations of spurious poles where Gram determinants
vanish. For technical details of the numerical implementation
we refer the reader to Ref.~\cite{Berger:2008sj}.

\subsection{Contribution from Infinity}
\label{sec:infinity}

The remaining rational contribution $\mathcal{R}_n^{\mbox{\tiny
large\,} z}$ is the boundary contribution in the contour integral
as $z \rightarrow \infty$. Although it is generically possible to
find complex continuations that have vanishing boundary
contributions, these shifts have in general additional
contributions that cannot be constructed recursively as in
eq.~(\ref{RationalRecursion}), that is, we have instead,
\begin{equation}
\mathcal{R}_n^P(0)  =  \mathcal{R}_n^{P\,\mbox{\scriptsize recursive}} +
 \mathcal{R}_n^{P\,\mbox{\scriptsize nonstd}} \, , \quad \mathcal{R}_n^{P\,\mbox{\scriptsize recursive}} \equiv
-\sum\limits_{{\rm poles}\ \beta}
\Res_{z=z_\beta}  {{\cal R}_n^{P} (z)\over z} \, . \label{nonstandard}
\end{equation}
These `non-standard' contributions
(labeled by the superscript ``nonstd'') arise in configurations where
two external momenta with the same helicity are on one side
of the partition and all other legs are on the other side.
The complex factorization properties of these configurations are not yet fully understood.
The sum over poles $\beta$ in eq.~(\ref{nonstandard}) is \emph{only} over the channels
that factorize, i.\,e. that do not display the aforementioned problematic behavior.
Conversely, it is possible to find shifts that avoid these `non-standard channels',
however, at the price of reintroducing a boundary contribution.

The solution to this problem, developed in Ref.~\cite{Berger:2006ci}, is to use
two independent complex continuations to determine the boundary contribution.
Let us denote the primary shift by $\Shift{j}{l}$ and the auxiliary shift
by $\Shift{a}{b}$ in the notation of eq.~(\ref{SpinorShift}). We then have
two relations for the same amplitude, analogous to eq.~(\ref{totalamp}),
\begin{eqnarray}
A_n^{\oneloop}(0)\! & = & \!\InfPart{\Shift{j}{l}}{A_n^{\oneloop}} + C_n(0)  - \InfPart{\Shift{j}{l}}{C_n}
 + \mathcal{R}_n^{P\,\mbox{\scriptsize recursive}, \,\Shift{j}{l}}   + \mathcal{R}_n^{S\,\Shift{j}{l}}(0)  \, ,
 \label{primaryrec} \,\,  \\
A_n^{\oneloop}(0)\! & = & \! C_n(0) - \InfPart{\Shift{a}{b}}{C_n}
+ \mathcal{R}_n^{P\,\mbox{\scriptsize recursive}, \,\Shift{a}{b}} +
\mathcal{R}_n^{P\,\,\mbox{\scriptsize nonstd}, \Shift{a}{b}} +  \mathcal{R}_n^{S\,\Shift{a}{b}}(0)
 \label{auxiliaryrec} \,  . \,\,\,\,\,\,\,\,\,\,\,
\end{eqnarray}
We have indicated with additional superscripts which shift has
been employed. $\mathcal{R}_n^S$ is evaluated according to
eq.~(\ref{spuriousfinal}). We have also used that,
\begin{equation}
\mathcal{R}_n^{\mbox{\scriptsize large\,} \tiny z} = \InfPart{\Shift{j}{l}}{A_n^{\oneloop}} - \InfPart{\Shift{j}{l}}{C_n} \, ,
\end{equation}
where $ \InfPart{\Shift{j}{l}}{A}$ is the unknown large-$z$
behavior of the full amplitude found from a Laurent expansion of
$A_n^{\oneloop}(z)$ around $z = \infty$ with the shift
$\Shift{j}{l}$, and similarly for $C_n(z)$.
Eqs.~(\ref{primaryrec}) and (\ref{auxiliaryrec}) thus both contain
unknown terms. If we now apply the primary shift $\Shift{j}{l}$ to
the auxiliary recursion (\ref{auxiliaryrec}), and take the limit
$z \rightarrow \infty$, we can extract the large-$z$ behavior of
the primary shift,
\begin{equation}
 \InfPart{\Shift{j}{l}}{A_n^{\oneloop}}  =
 \InfPart{\Shift{j}{l}}{C_n} -
\InfPart{\Shift{j}{l}}{\bigg(\InfPart{\Shift{a}{b}}{C_n}\bigg)}
                    + \InfPart{\Shift{j}{l}}{\bigg({\cal R}^{P\,\mbox{\scriptsize recursive} \, \Shift{a}{b}}_n  \bigg)}
                    + \InfPart{\Shift{j}{l}}{\bigg(\mathcal{R}_n^{S\,\Shift{a}{b}}(0_{\Shift{a}{b}})\bigg)} \, ,
\label{infpartfinal}
\end{equation}
where now all terms on the right-hand side are either known or
recursively constructible, \emph{if}
\begin{equation}
 \InfPart{\Shift{j}{l}}{\bigg( \mathcal{R}_n^{P\,\,\mbox{\scriptsize nonstd}, \Shift{a}{b}}\bigg)}
 = 0 \, . \label{vanish}
\end{equation}

Putting everything together, we find the full amplitude from eq.~(\ref{primaryrec}),
with (\ref{RationalRecursion}),
(\ref{spuriousfinal}), and (\ref{infpartfinal}), and the cut terms constructed as
described in the previous section.

\section{Conclusions and Outlook}

In this review we have presented an overview of recent developments in the calculation of multi-parton
scattering amplitudes at the one-loop level. These developments are based on on-shell techniques
that make efficient use of the physical properties of the hard scattering, such as unitarity and
factorization. The basic ingredients in these new approaches are on-shell tree-level or lower-point
one-loop amplitudes instead of Feynman diagrams, thus sidestepping many of the complications
associated with the use of Feynman diagrams.

Furthermore, these new techniques allow
for efficient algorithmic implementation and hence the construction
of efficient, numerically stable, and fast computer codes, such as \texttt{BlackHat}~\cite{Berger:2008sj},
 \texttt{CutTools/OneLOop}~\cite{Ossola:2007ax,vanHameren:2009dr}, \texttt{Rocket}~\cite{Ellis:2008qc}, and others~\cite{othercodes}.
With these new techniques and computer tools a flurry of results relevant for the LHC has recently been
computed~\cite{Berger:2009zg,KeithEllis:2009bu,Melnikov:2009dn,Ossola:2007bb,Binoth:2008kt,Bevilacqua:2009zn},
and we expect further rapid progress in the near future.

Nevertheless, much work remains to be done
to bring one-loop calculations to the same level of automatization as tree-level computations,
 ideally
starting from a Lagrangian and producing complete events including
parton shower and hadronization corrections. Further open issues
include, for example, a better understanding of the complex
factorization  properties of one-loop amplitudes and the
generalization of the new techniques to higher loops in
nonsupersymmetric theories.

In summary, the last few years have seen an unprecedented progress
in the development of techniques for the computation of
multi-parton one-loop scattering amplitudes which are an essential
ingredient in precision calculations for the LHC. These new
methods have also
 been used to study the higher-loop structure of ${\cal N} = 4$
supersymmetric Yang-Mills theory and ${\cal N} = 8$
supergravity~\cite{sugra}.  Their basic ingredients are unitarity,
factorization and complex analysis, properties that are quite
generic. It is thus not inconceivable that these new techniques
will find further application beyond those presented or referenced
in this review.

\section*{Acknowledgments}

We would like to thank Zvi Bern, Lance Dixon, Fernando Febres
Cordero, Tanju Gleisberg, Harald Ita, David Kosower, and Daniel
Ma\^{i}tre for fruitful collaboration. We also thank Zvi Bern and
Lance Dixon for valuable comments. CFB thanks the Aspen Center for
Physics for hospitality. This work is supported in part by funds
provided by the U.S. Department of Energy under cooperative
research agreement DE-FC02-94ER40818.

\end{document}